\documentclass{aa}
\usepackage{graphics,aanda,amssymb,rotating,psfrag,amsmath,epsfig,array}
\usepackage{color}
\begin{document}

\title{ISO observations of the interacting galaxy Markarian 297
\footnote{Based on observations with ISO, an ESA project with instruments
funded by ESA Member States (especially the PI countries: France,
Germany, the Netherlands and the United Kingdom) with the participation
of ISAS and NASA}  
}
\subtitle{with the powerful supernova remnant 1982aa}

\titlerunning{ISO observations of the interacting galaxy Markarian 297}

\author{L.\,Metcalfe\inst{1,2}   \and
        B.\,O'Halloran\inst{3,4} \and 
        B.\,McBreen\inst{3}      \and  
        M.\,Delaney\inst{3}      \and
        M.\,Burgdorf\inst{5}     \and 
        K.\,Leech\inst{2}        \and
        P.\,Barr\inst{6}         \and
        J.\,Clavel\inst{6}       \and
        D.\,Coia\inst{1,3}       \and
        L.\,Hanlon\inst{3}       \and
        P.\,Gallais\inst{8}      \and
        R.\,Laureijs\inst{6}     \and
        N.\,Smith\inst{9}        }

\offprints{leo.metcalfe@sciops.esa.int}

\institute{XMM-Newton Science Operations Centre, European Space Agency, Villafranca del Castillo, P.O. Box 50727, 28080 Madrid, Spain. \and 
           ISO Data Centre, European Space Agency, Villafranca del Castillo, P.O. Box 50727, 28080 Madrid, Spain. \and 
           Physics Department, University College Dublin, Dublin 4, Ireland. \and
           Dunsink Observatory, Castleknock, Dublin 15, Ireland.             \and
           SIRTF Science Center, California Institute of Technology, 220-6, Pasadena, CA 91125, U.S.A. \and
           Research and Scientific Support Department, European Space Agency, ESTEC, Keplerlaan 1, 2200 AG Noordwijk ZH, The Netherlands. \and
           CEA Saclay/Service d'Astrophysique, Orme des Merisiers, F-91191 Gif-sur-Yvette Cedex, France. \and
           Cork Institute of Technology, Cork, Ireland.                     \\}

\date{Received / Accepted }

\abstract{ Markarian (Mkn) 297 is a complex system comprised of two
interacting  galaxies that has been modelled with a variety of
scenarios.  Observations of this system were made with the Infrared
Space Observatory (ISO) using the ISOCAM, ISOPHOT and LWS
instruments. ISOCAM  maps  at  6.7\,$\mu$m, 7.7\,$\mu$m,
12\,$\mu$m  and  14.3\,$\mu$m  are  presented which, together with
PHT-S spectrometry of the central interacting region, probe the
dust obscured star formation and the properties of the organic
dust. The ISOCAM observations reveal that the strongest emission in
the four bands is at a location completely unremarkable at visible
and near-IR (e.g. 2MASS) wavelengths, and does not coincide with the
nuclear region of either colliding galaxy. This striking characteristic 
has also been observed in the overlap
region of the colliding galaxies in the Antennae (NGC 4038, 4039),
the intragroup region of Stephan's Quintet, and in IC 694 in the
interacting system Arp 299, and again underlines the importance
of infrared observations in understanding star formation in
colliding/merging systems.  At 15\,$\mu$m, the hidden source in 
Mkn 297 is, respectively, 14.6 and 3.8 times more luminous than the hidden 
sources in the Antennae (NGC\,4038/4039) and Stephan's Quintet. 
Numerical  simulations of the Mkn\,297 system indicate that a co-planar
radial penetration between two disk galaxies yielded the observed
wing formation in the system about 1.5\,$\times$ 10$^{8}$ years
after the collision.  A complex emission pattern  with knots and
ridges of emission was detected with ISOCAM.  The 7.7\,$\mu$m map
predominantly shows the galaxy in emission from the 7.7\,$\mu$m 
feature attributed to PAHs (Polycyclic Aromatic  Hydrocarbons). The
14.3/7.7\,$\mu$m ratio is greater than unity over most of the
galaxy, implying widespread  strong star formation.  Strong
emission features were detected in the ISOPHOT spectrum, while
[O\,I], [O\,III] and [C\,II] emission lines were seen with LWS.
Using data from the three instruments, luminosities and masses for two
dust components were determined. The total infrared luminosity is
approximately 10\(^{11}\)\,L$_{\odot}$, which (marginally) classifies 
the system as a luminous infrared galaxy (LIRG).   A supernova that 
exploded in 1979 (SN\,1982aa) gave rise 
to one of the most powerful known radio remnants which falls close to 
the  strongest mid-infrared source and is identified with star forming
region 14 in the optical.  This supernova explosion may have been accompanied
by a gamma-ray burst (GRB), consistent with the idea that GRBs are
associated with supernovae in star forming regions, and a search
for a GRB  consistent with the direction to Mkn\,297, in satellite 
data  from July to December 1979, is recommended. }


\maketitle

\keywords{Galaxies: general -  Mkn\,297 -  Galaxies: interactions -
galaxies: starburst - (ISM:) dust, extinction - Infrared: galaxies}

\section{Introduction}

Mkn\,297 (also known as NGC\,6052, Arp\,209 and UGC\,10182) has been the subject
of many investigations because of its peculiar   
morphology \cite{weed:1972}.   It has been referred to as a typical
clumpy irregular galaxy  \cite{tt:1987} and a giant HII 
region galaxy  \cite{thu:1983,thuma:1981}.  It has a  
complex optical structure  consisting of forty three identified bright  
optical knots within a common envelope \cite{hcm:1987}. 
The galaxy has no obvious companion and appears to be isolated.  
However, it has two distinct components revealed through optical
spectroscopy \cite{duf:1976,bur:1988}, CO \cite{sls:1993} and HI
line emission \cite{sthw:1990}, and consequently the system has
been variously modelled as a colliding system consisting of  two 
late-type spiral galaxies \cite{ad:1979}; the collision of a
spiral galaxy with an irregular galaxy \cite{bur:1988}; or more
recently, using numerical N-body simulations, as the collision of 
two disk galaxies \cite{tn:1991}. Throughout this paper we will refer to
the component identified as a spiral by Burenkov \cite*{bur:1988} as galaxy A and
the other component as galaxy B.

The two components have a velocity separation of about
200\,km\,s$^{-1}$  and modelling reveals that their closest 
approach  occurred  about 1.5\,$\times$\,10$^{8}$  years  ago
\cite{tn:1991}.   Mkn\,297  has  similar infrared  properties  to 
the Antennae  Galaxies  (NGC\,4038/4039)  and,  as  in the
Antennae, the strongest infrared  source does not coincide with the
nuclei      of     the      interacting      galaxies
\cite{vma:1996,krl:1996,mira:1998} nor, indeed, with any other feature
notable at visual or near-IR wavelengths (see Figs.\,1, 2 and 3 below). 
Heeschen et al.  \cite*{hhy:1983}
have found a peculiar compact and variable radio source, Mkn\,297A,
lying several arcseconds  north   of  the nuclei of the 
interacting  galaxies (see Fig.\,8).   They interpreted  the compact  source as  a
complex  of  supernova remnants associated with a region of 
vigorous star formation.  Yin \& Heeschen \cite*{yh:1991} and Deeg 
et al. \cite*{dbd:1993} interpret Mkn\,297A as a single luminous
radio supernova which exploded in July 1979. Mkn\,297  was  observed
by  ROSAT, and has an X-ray luminosity L$_{X}$ = 
2 x 10\(^{41}\)\ ergs s$^{-1}$ \cite{hen:1997} with ROSAT PSPC
around 1keV, a value that is comparable to the cumulative flux 
of the Antennae observed with Chandra at similar energy \cite{zez:2002}. 

The    capabilities     of    the    Infrared     Space    Observatory
(ISO) \cite{ksa:1996} permitted observations  of Mkn\,297
across a wide range of IR wavelengths with  three  complementary  instruments. 
The observations  and  data
reduction  techniques are  presented  in Sec.\,2.   The results  are
presented in Sec.\,3 and discussed in Sec.\,4. A value of H$_{0}$ =
72\,km\,s\,$^{-1}$\,Mpc$^{-1}$ has been  adopted and yields a distance
of 66\,Mpc to Mkn\,297. 

\section{Observations and Data Reduction}

\begin{table*} 

\caption{The log of the ISO observations of Mkn\,297.  The
nine columns list the TDT number (a unique identifier for an ISO
observation), the AOT number (which identifies the observing mode
used), the filter label, the wavelength range ($\Delta\lambda$),
the reference wavelength of the filter where appropriate, the duration of 
the measurement (including both the on- and off-source measurements), the 
field-of-view of the instrument in the configuration employed, the positional 
offset of the background measurement(s) with respect to the target reference 
position, and any additional notes.}
\begin{center} 
\begin{tabular}{cclcccccl}\hline 
\noalign{\smallskip}
TDT\#$^{a}$ & AOT\# & Filter  & $\Delta\lambda$  &  $\lambda_\mathrm{ref}$ & $\Delta$t & F.O.V.                            & \ Chop                & Notes         \\
          &         &         & ($\mu$m)         & ($\mu$m)                &  (sec.)   & \ $^{\prime\prime}$               & \ \ $^{\prime\prime}$ &               \\
          &         &         &                  &                         &           &                                   &                       &               \\
\noalign{\smallskip}
\hline
\noalign{\smallskip}
47400671  &  CAM03  &    LW2  &     5.0--8.5     &             6.7         &    588    &   48 (1.5$^{\prime\prime}$/pixel) &     90                & chop in Decl. \\
47400671  &  CAM03  &    LW6  &     7.0--8.5     &             7.7         &    588    &   48                              &     90                & chop in Decl. \\
47400671  &  CAM03  &    LW3  &    12.0--18.0    &            14.3         &    588    &   48                              &     90                & chop in Decl. \\
47400671  &  CAM03  &    LW10 &     8.0--15.0    &            12.0         &    588    &   48                              &     90                & chop in Decl. \\
09101068  &  CAM01  &    LW3  &    12.0--18.0    &            14.3         &   3794    &  480 (6$^{\prime\prime}$/pixel)   &                       & raster$^{b}$  \\
47400567  &  PHT40  &   SS/SL &     2.5--11.6    &                         &   1024    &   25                              &  +/-150               & triang. chop$^{c}$ \\
62702069  &  LWS01  &         &    43.0--190.0   &                         &   1124    &  101                              &                       & on-source     \\
62702070  &  LWS01  &         &    43.0--190.0   &                         &   1124    &  101                              & see below$^{a}$       & off-source    \\
\noalign{\smallskip} \hline
\noalign{\smallskip} 
\multicolumn{9}{l}{  (a) All observations refer to target position R.A. 16\(^{h}\) 05\(^{m}\) 13.2\(^{s}\) \& Dec. +20$^{\circ}$ 32$'$ 32.0$''$ } \\ 
\multicolumn{9}{l}{  except for the LWS background measurement made at 16\(^{h}\) 05\(^{m}\) 57.2\(^{s}\) +20$^{\circ}$ 26$'$ 23.7$''$. }     \\ 
\multicolumn{9}{l}{  (b) TDT 09101068 was a $4 \times 4$ raster with 96$^{\prime\prime}$ step size. }               \\ 
\multicolumn{9}{l}{  (c) The term ``chop" has been used to refer to all off-target measurements, although only PHT } \\ 
\multicolumn{9}{l}{  employed a chopping device. For the other instruments the spacecraft pointing was offset.} \\ 
\noalign{\smallskip} 
\noalign{\smallskip}
\end{tabular} 
\end{center} 
\label{iso-log}
\end{table*}

The  ISO  observations were  obtained  using: (a) The mid-infrared 
camera ISOCAM \cite{iso-cesarsky-1996aa}, mainly in beam--switching
mode, though one larger--area raster observation has been used; (b)
the spectrometric channel (PHT-S) of the ISO photopolarimeter
(ISOPHOT) \cite{iso-lemke-1996aa} in triangular chopping mode and
(c) the medium-resolution  grating mode of the long wavelength
spectrometer LWS \cite{iso-clegg-1996aa} employing dedicated on- and
off-source measurements. The parameters of these observations are
listed in Table 1 and discussed in the following sections.

\subsection{ISOCAM}

A detailed technical description of the ISOCAM instrument and  its observing
modes can be found in \cite{iso-blommaert-2003}.  For the present
work the 1.5$''$ per pixel plate scale of the instrument  has been
used yielding a field  of view of 48$''$ $\times$ 48$''$.  As
explained in Section 2.4 an available lower-resolution
6$^{\prime\prime}$/pixel map has been used only to rule out the
presence of  very extended mid-IR emission from the system.
Observations were obtained in  four filters, with reference
wavelengths \cite{mon:1997}  at 6.7\,$\mu$m  (LW2), 7.7\,$\mu$m 
(LW6), 12\,$\mu$m (LW10) and  14.3\,$\mu$m (LW3).  Each filter 
observation consisted of two measurements,  one centred 1.5\arcmin\
away  from the target,  and the other centred on the target
coordinates. Each of these beams was observed  for 140 on-chip
integrations  of 2.1 seconds each, for  a total  dwell-time  per
filter of 588 seconds. The diameter of the ISO point spread function
(PSF) central maximum, in arcseconds,  at the first Airy minimum is
0.84\,$\times\,\lambda(\mu$m). The FWHM is about half that, and
Okumura (1998) obtained a value of 2.3$''$ at 6.7\,$\mu$m and
4.6$''$ at 14.3\,$\mu$m for the PSF FWHM in the 1.5$''$ per pixel
configuration.

All data processing used the CAM Interactive Analysis
(CIA) software \cite{iso-ott-1997asp,iso-ott-2001asp,iso-delaney-2002esa}, 
as follows: (i) Dark
subtraction was performed using a dark model with correction for slow
drift of the dark--current throughout the mission; (ii) glitch effects
due to cosmic rays were removed following the method described in 
Starck et al. \cite*{star:1998}; 
(iii) transient correction for signal attenuation due to the lag
in  the detector  response was  performed by  the method  described by
Abergel et  al. \cite*{aab:2000}; (iv) the maps were  flat-fielded using 
library flat-fields; (v) pixels  affected by glitch residuals and other
persistent effects  were manually suppressed; (vi) the on-target and
off-target beams were subtracted and (vii) the images were deconvolved
using a multi-resolution transform method \cite{smb:1998}.

Considerable   care   was   needed   in  the   preparation   of  
the 14.3/6.7\,$\mu$m  ratio  map.   Apart from  deconvolution 
and  background subtraction \cite{ohalloran2002}, it  was
necessary to eliminate  any misalignment between the  14.3 and
6.7\,$\mu$m  images due  mainly to  some jitter\footnote{Some 
``play'' was designed into the CAM cryo-mechanisms
to avoid mechanism seizure at LHe temperatures.}  in the
positioning  of   the  CAM  wheels,  leading  to   small  offsets  in
astrometry.  This was done by choosing a suitable reference source in both
images and locating its position  using a gaussian fit.  
The coordinates of  the reference object in each image were  established 
at sub-pixel accuracy by the fit and the  vector distance  between the 
reference source positions in successive images was determined, establishing 
the  extent of misalignment. A PSF  was
chosen, from a library of model PSFs, offset from the  centre by
the negative of the distance vector.  This  PSF was used to
deconvolve and align the misaligned images.  The reference image was then
deconvolved  with a  centred PSF so the resultant images were
aligned and had the same beam. Noise spikes in the ratio map, due to the  
division of background pixels having noise-related low apparent signal 
values were suppressed by applying the same positive threshold to the background
pixels in both images before division.  Note that photometry of sources 
was performed on the un-deconvolved maps using the CIA  routine XPHOT, employing 
aperture photometry scaled for the PSF.

The durations of the observations in Table 1 include  
instrumental,  but   not  spacecraft,
overheads.  The flux densities quoted in Table 2 for four source regions
assume a spectrum having $\lambda F_{\lambda} = const.$, and have a
photometric accuracy of about $\pm$\,15\%.

\subsection{PHT-S}

PHT-S consists of a dual grating spectrometer with  a resolving
power of 90  \cite{iso-laureijs-2003}.   Band   SS  covers   the 
range 2.5\,-4.9\,$\mu$m,     while    band     SL    covers    
the    range 5.8\,-\,11.6\,$\mu$m.    The  PHT-S   spectrum  of  
the  core region of the Mkn\,297 system was  obtained 
by  operating  the  $24''\times24''$ aperture of PHT-S in
triangular chopping mode using the ISOPHOT focal--plane chopper.  
In this mode, the satellite pointed to the given target coordinates
centred between two off-source-positions, and the chopper moved
alternately from the target to the first off-source position, back 
to the target and
then to the second off-source position. Consequently, 50\% of the exposure
time is on target and 50\% is distributed equally over the two
off-source positions.
In the on-target phase of the chop the PHT-S aperture was centred on the
centre of the ISOCAM pointing, a position about 3 arcseconds
East of the centre of the 15\,$\mu$m image presented in Fig.\,1, as indicated
in the figure.   The $24''\times24''$ PHT-S aperture
covered completely ISOCAM source regions 1, 2 and 4, and  vignetted
slightly the other regions. (See Sections 2.4 and 3.3 below.)

PHT data processing was performed using the ISOPHOT
Interactive  Analysis  (PIA)  system, version  8.1
\cite{iso-gabriel-2002esa}. Data reduction consisted primarily of
the removal of instrumental effects such  as ionising particle impacts
which  result in a spurious increase in two or more consecutive read-out
voltage values.   The disturbance is usually very short and  the
slope of  the integration ramp after a glitch is  similar  to the
slope before it.   Further reduction of the spectrum  was performed by
manual deglitching, where the slopes of the ramps were considered for a
given  PHT-S channel,  and two  possible corrections  were applicable.
First a  strong startup transient was searched for (Acosta-Pulido, Gabriel \& 
Casta\~{n}eda 2000) and, if present, suppressed 
by  deleting  either the initial 64 or 128 seconds of integration.   
The remaining signals were inspected for large deviations from the mean, indicating
strong cosmic ray hits which could not be removed by  the automatic deglitching.  
These anomalous signals
were  suppressed manually by masking data affected by the glitch.  The mean 
differential signal (i.e. the difference of the on-source and off-source 
chopper beams) for each of the 128 pixels of PHT-S was derived. To absolutely
calibrate the resulting spectrum it was scaled to match the signal level
of the ISOPHOT-S Highly Processed Data Products for this target which can
be found in the ISO archive \cite{ricklaas:2002} associated with the Mkn\,297 
archive entry for this observation. The HPDP are somewhat noisier than our
PIA produced product, but benefit from being part of a systematic reprocessing
of a large sample of PHT-S spectra, and we consider that they are likely to 
have a better overall calibration.  Relative (from point to point in the 
spectrum) and absolute (overall normalisation of the spectrum to flux standards)
spectrometric accuracies of better than 10\% are indicated.
The results were plotted to obtain the spectrum for the region of Mkn\,297 in 
the PHT-S aperture.
Flux values derived from this spectrum have to be further scaled, as 
described in Sec.\,2.4 below, to yield values more appropriate to the 
galaxy as a whole.


\subsection{Long Wavelength Spectrometer (LWS)}

A detailed technical description of the LWS instrument and  its observing
modes can be found in \cite{iso-gry-2003esa}. A LWS spectrum  
of Mkn\,297 spanning the range from 45 to 180\,$\mu$m was obtained 
with the LWS aperture centred on the ISOCAM map, at a position about 
3$''$ East of  the nucleus of galaxy A, as indicated by
the filled square in Fig.\,1. The beam of LWS was
slightly elliptical and its FWHM varied between  65$''$ and 85$''$,
depending on wavelength and direction (Swinyard et al. 1998). It was 
assumed that the  source  was completely included  in the beam of 
the LWS instrument, so that no extended-source correction was necessary. 

Because of an expected significant background signal in the LWS 
aperture due to combined Zodiacal and Cirrus emission in this sky 
region, a dedicated background measurement was made with the LWS
aperture centred about 12
arcminutes away from the on-source measurement (at a position determined
to be of similar background based on the IRAS maps), and the resulting
background was subtracted from the on-target case.

The  grating was scanned 6 times over  the entire  wavelength
range.   The spectral sampling was set to give 4
samples per resolution element in each of the scans. 

We made use of the LWS Highly Processed Data Products (HPDP) for this target, 
available from the ISO archive (Lloyd et al. 2003).
Because of
the small flux from the source and the resulting  poor
signal-to-noise ratio  of the  LWS continuum spectrum, the data were 
rebinned to one point for each of the 10 LWS detectors, employing a
scan-averaging method described by Sidher et~al. (2000), yielding
10 samples of the continuum spanning the LWS spectral range and plotted 
in Fig.\,7.


\subsection{Handling of different instrument fields-of-view}

The rather different apertures, or fields-of-view, of the
instruments  warrant comment.  Observatory pipeline products from the 
low-resolution ISOCAM 6$^{\prime\prime}$/pixel raster observation 
of ISO TDT 09101068 have been used to establish that substantially less
than 10\% of the total galaxy MIR light falls outside of the
1.5$^{\prime\prime}$/pixel, 48$^{\prime\prime}$ field-of-view 
upon which the results reported here
rely. The much larger LWS aperture (see Section 2.3 above) is
assumed  to have covered the full extent of the galaxy, even at the
longer wavelengths addressed by LWS. The PHT-S aperture is a square
24$^{\prime\prime}$ on a side, and was placed as shown in Figure
1. It certainly missed some of the integrated galaxy signal in the
PHT-S wavelength range, although it included most of the knots of
mid-infrared emission found in the ISOCAM map. But the PHT-S aperture
has a vignetted profile (Laureijs et al. 2003). So some scaling has
to be applied to PHT-S signals to adapt them to the signals that
would have been recorded in corresponding spectral ranges over the full 
area of the ISOCAM footprint. To determine the scaling factor the 
PHT-S spectrum was used to
synthesise a 7.7\,$\mu$m (LW6) ISOCAM bandpass measurement by
integrating the CAM filter  profile over the PHT-S spectrum. The
result was compared with the CAM 7.7 micron  global galaxy
photometry and found to differ by a factor of 1.46. PAH feature
strengths derived from the PHT-S spectrum were scaled by this factor
(Sec.\,3.3).

\section{Results}

\subsection{A hidden source}

\color{black}

\begin{figure*}
\epsfig{file=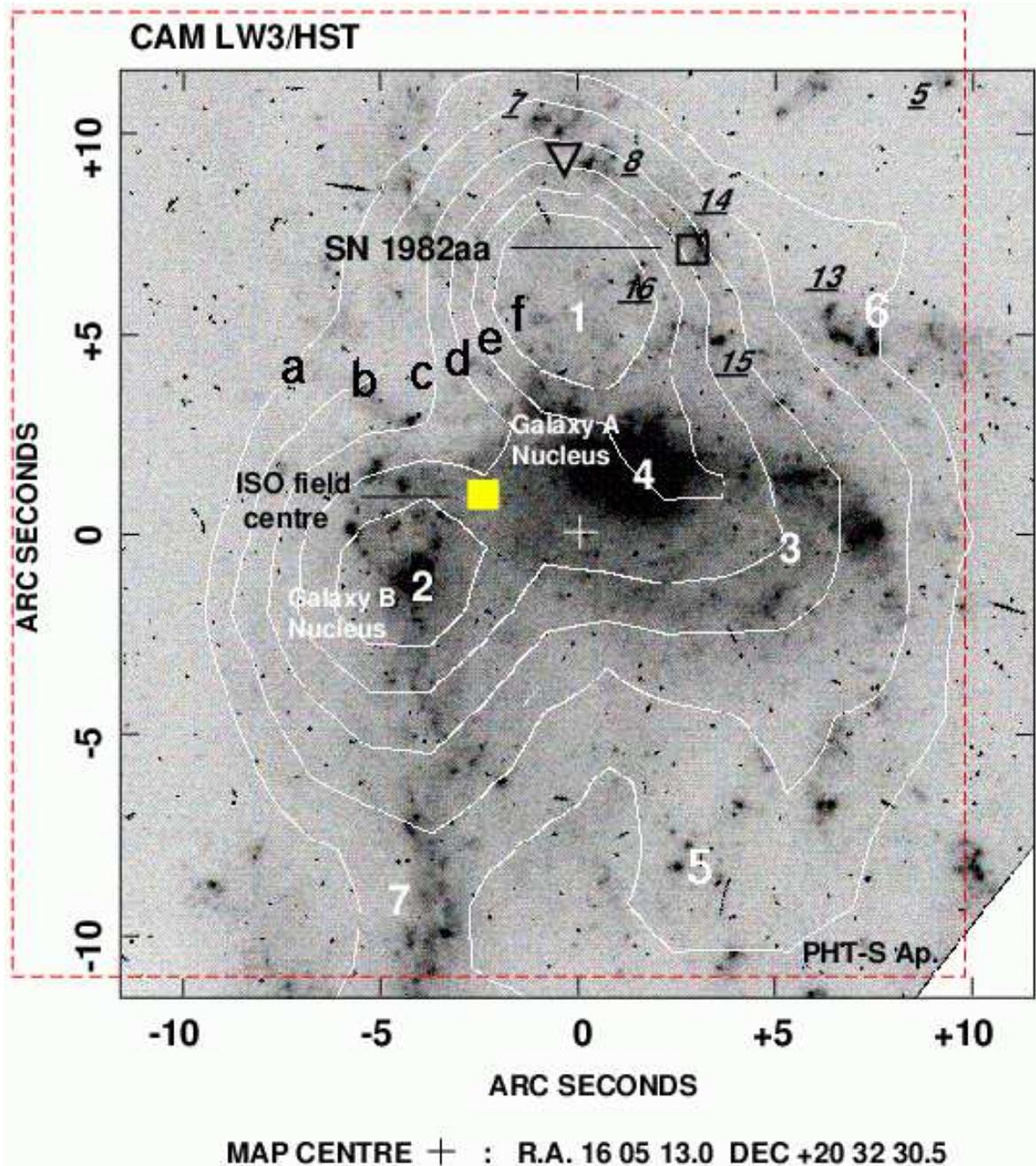,width=\textwidth,clip=,bb=17 17 593 773}
\vskip -5.0cm
\caption{A deconvolved  ISOCAM  14.3\,$\mu$m  contour  map  of  Mkn\,297
overlaid on a HST image (Filter: F555W, Centre: 5407\,$\AA$, Bandwidth: 1236\,$\AA$).
The contour levels, in the order a to f, are
as follows (in  units of mJy/arcsec$^{2}$): 0.7, 1.0,  1.5, 2.0, 2.5
and 3.0. The seven regions identified by white numbers correspond to the positions of 
well-defined structures or extensions of emission seen in one or more of the bandpasses 
presented in Fig.\,2.  Small black underlined numbers identify some H$_{\alpha}$
emission knots as designated by Hecquet et al.\,(1987), and help relate 
this map to earlier published data. Source 4 coincides approximately with the nuclear
region of galaxy A  and source 2 with  the nuclear region of  galaxy B, oriented
north-south. However the strongest source, no. 1, does not
correspond to either nucleus. The open square denotes the
position of one of the brightest known radio supernovae, SN\,1982aa (Mkn\,297A) and
corresponds to H$_{\alpha}$ source 14.
The position of a knot of 2MASS K-band emission  is indicated by the small black 
triangle (see also Fig.\,3). The filled light--coloured square designates the position of the
centre of the PHT-S aperture, and the  large dashed square on which
it is centred indicates the size of the PHT-S aperture. The LWS
aperture had the same centre, but was much larger than the region
shown here.}
\end{figure*}

The deconvolved 14.3\,$\mu$m map of Mkn\,297 overlaid on a HST
image (filter F555W at 5407\,$\AA$) is presented in Fig.\,1, while
the deconvolved maps at 6.7, 7.7, 12 and  14.3\,$\mu$m are
presented in Figs.\,2i\,--\,2iv.  The figures reveal a  complex 
emission pattern  of overlapping sources  and seven source regions 
are numbered  on the maps.   The integrated CAM  flux densities for the system
are presented in Table 2  along with estimated  flux densities  from
the four brightest MIR clumps. Knots that appear well defined in one bandpass
may not be so sharp in another (see Fig.\,2). Representative flux-density values have
been derived by performing aperture photometry about the locations
numbered in 1 to 4 in Figs.\,1 and 2 and scaling each result to  the PSF that would
deliver the same signal in the aperture used. Where no well-defined
point-source appears in some bandpass these ``point-source''  flux 
densities nevertheless serve to characterise relative brightness 
over the map.  Source 4
coincides  approximately with  the nuclear region of galaxy A and
source 2 with the nuclear region of galaxy B.  The strongest 
source, labeled 1 on all the CAM maps, does  not coincide with 
the nucleus  of either galaxy  but is displaced by  about 2 kpc
from  the nucleus  of galaxy A and peaks over a region devoid of
optical emission knots.  (We stress, by the way, that this source is clearly 
distinguished also in the un-deconvolved maps.) Hecquet et 
al.\,\cite*{hcm:1987} reported numerous optical
emission knots in Mkn\,297, and knots 7 and 8 in their numbering
scheme have the second and third highest values of B -- R (1.8 and 2.3 respectively) 
for the system and are within a few
arcseconds of the mid-infrared peak.  Knot 14 is the site of the variable radio
source reported by Heeschen et al.\,(1983). These optical knots may represent the visible
portions of a much vaster obscured starburst. A similar
result has been obtained 
\cite{vma:1996,mira:1998} for the interacting galaxies 
NGC\,4038/4039 (The Antennae) where the strongest mid-infrared source 
does not coincide with either of the nuclei of the
constituent galaxies.

\begin{table}
\caption{The ISOCAM flux densities for MIR clumps in Mkn\,297. 
The photometry was performed before deconvolution and is
accurate to about $\pm$\,15\%.}
\begin{center}
\begin{tabular}[h]{rccccc}
\hline\noalign{\smallskip}  
   $\lambda$  &  Total  Flux  & \multicolumn{4}{c}{Source    Flux Densities} \\    
              &    Density    &            &            &           &        \\
    ($\mu$m)  &     (mJy)     & \multicolumn{4}{c}{(mJy)}                    \\ 
              &               &     1      &     2      &     3     &    4   \\
\noalign{\smallskip}  
\hline\noalign{\smallskip}  
  6.7         &      141      &   19.5     &   16.0     &    14.0   &  19.0  \\
  7.7         &      246      &   30.5     &   19.5     &    19.0   &  29.5  \\
 12.0         &      269      &   42.5     &   28.0     &    24.5   &  31.5  \\
 14.3         &      289      &   66.5     &   48.5     &    30.5   &  43.5  \\
\noalign{\smallskip}\hline
\end{tabular}
\end{center}
\end{table}

\begin{figure*}
\epsfig{file=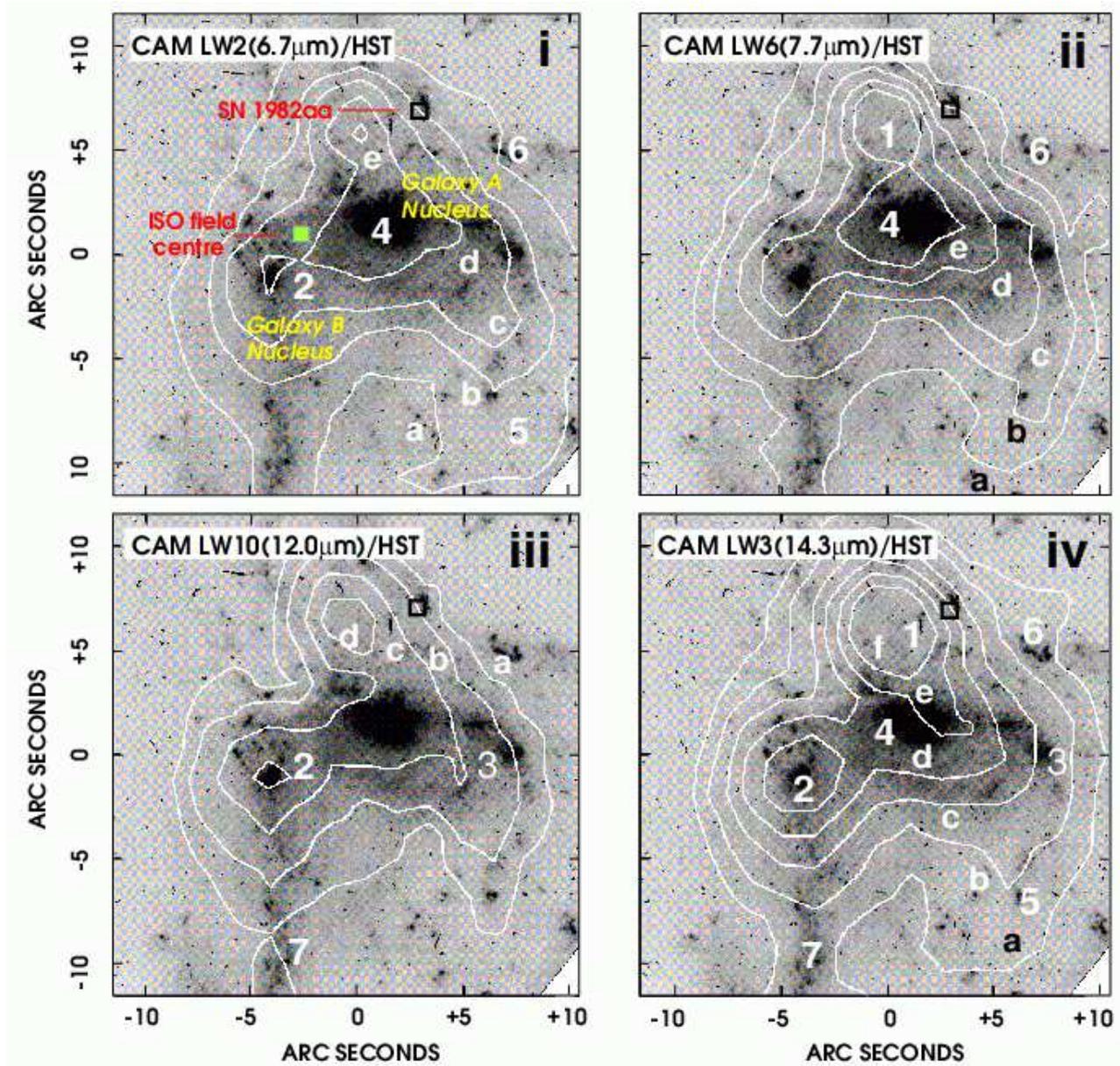,width=\textwidth,clip=,bb=17 17 593 773}
\vskip -3cm
\caption{Deconvolved ISOCAM  contour maps overlaid on the HST image: 
  i) 6.7$\mu$m, ii) 7.7$\mu$m, iii) 12$\mu$m and iv) 14.3\,$\mu$m maps are presented.  
  Seven regions of emission
  are  identified (the numbers running from 1 to 7).  
  Sources 2  and 4   coincide  with the nuclear  regions  of  the
  constituent interacting galaxies.  The  contour 
  levels in the sequence a to f  are as follows
  (in units  of mJy/arcsec\(^{2}\))~:~i) 0.4,  0.7, 1.0, 1.5  and 2.0,
  ii) 0.7, 1.0,  1.5, 2.0 and 2.5, iii) 0.7, 1.0, 1.5 and 2.0, iv)
  0.7, 1.0, 1.5, 2.0, 2.5 and 3.0.}
\end{figure*}

\begin{table}
\caption{IRAS flux densities for Mkn\,297 (from Sanders et al. (2003)).  The four columns
list the filter used, the  wavelength range, the  reference wavelength for the filter
and the flux density measured.}
\begin{center}
\begin{tabular}{lccc}
\hline\noalign{\smallskip}     
     Filter       &  $\Delta\lambda$ &  $\lambda_\mathrm{ref}$ &    Total  Flux Density \\
                  &   ($\mu$m)       &  ($\mu$m)               &        (mJy)           \\
\hline\noalign{\smallskip} 
IRAS  12\,$\mu$m  &  8.5 -  15       &   12                    &   280\,$\pm$\,22       \\
IRAS  25\,$\mu$m  &  19 -  30        &   25                    &   830\,$\pm$\,27       \\  
IRAS  60\,$\mu$m  &  40 -  80        &   60                    &  6790\,$\pm$\,45       \\
IRAS 100\,$\mu$m  &  83 - 120        &  100                    & 10570\,$\pm$\,340      \\
\noalign{\smallskip}\hline
\end{tabular}
\end{center}
\end{table}

In order to compare the ISOCAM map with previously published
results at other wavelengths
\cite{hhy:1983,hcm:1987,lls:1992,dbd:1993,ddb:1997} it is 
necessary to consider some inconsistencies in the published
astrometry of  features in the Mkn\,297 system. The first three of the 
above--referenced papers use astrometry reported by Heeschen et al.\,\cite*{hhy:1983}, 
derived ultimately from glass
copies of original photographic plates and attributed in 
Hecquet et al.\,\cite*{hcm:1987}
to a private  communication from Casini in 1980.  However, a careful comparison
of the Hecquet et al.\,\cite*{hcm:1987} astrometry (their Table 1 and Fig.\,2b) with 
2MASS and HST images reveals a roughly 6$^{\prime\prime}$ offset of their coordinates to the North West.
The resulting  positions for optical knots in Mkn\,297 differ
systematically from positions derived by Deeg et al.
\cite*{dbd:1993} based on independent CCD photometry. Similarly,
the radio supernova reported by  Heeschen et al.\,\cite*{hhy:1983}
and Yin\,(1994), which from Figs.\,1, 2 and 8 of  this paper
clearly corresponds to H$_{\alpha}$ source 14 of Hecquet et
al.\,\cite*{hcm:1987}, only falls at that location subject to the
above adjustment of the astrometry of Hecquet et
al.\,\cite*{hcm:1987}.  The astrometry of Casini, Heeschen et
al.\,\cite*{hhy:1983}, Yin et al.\,(1994) and Deeg et al.\,(1993)
appears to be mutually consistent with all recent imaging data.
In fact, Lonsdale et al.\,\cite*{lls:1992} noted the same discrepancy and
refer to a private communication from Hecquet and Coupinot\,(1991)
in which they concur. 

\begin{figure}
\resizebox{\columnwidth}{!}{\includegraphics*{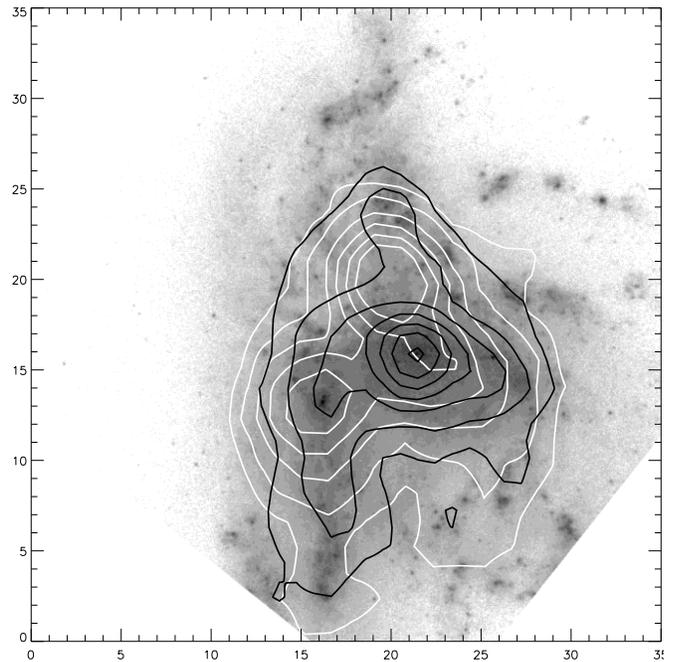}} 
\caption{An overlay of the ISO 15\,$\mu$m (white contours) and 2MASS K-band (black 
contours) on 
the HST image of the Mkn\,297 galaxy. The HST image and ISO contours are as in Figure 1.
The 2MASS contours are equispaced at levels 350, 355, 360, 365, 370, 375 and 380 
detector units. 
(We have not absolutely calibrated the 2MASS data, being only interested here 
 in spatial information.)
}
\end{figure}

Fig.\,1 therefore adopts
the 2MASS astrometry referenced to ISOCAM source 4, assumed to
correspond to the  brightest knot in the 2MASS K-band data. This
places ISO source 2 over the nucleus of galaxy B.  Fig.\,3 presents an 
overlay of the ISO 15\,$\mu$m (white contours) and 2MASS K-band (black contours)
on the HST image of Mkn\,297 used in Fig.\,1. It is clearly seen that 
ISO source 1 is not evident in the 2MASS map. (See also the J, H and K 
maps published by Smith et al.\,\cite*{shh:1996}.)
The K-band data of Cair\'{o}s et al. \cite*{cai:2003} also show a source
roughly 4 arcseconds North of our source 1 and consistent with the 
location of the 2MASS source and the optical knots.

Within the accuracy of the various sources of positional
information used to construct Fig.\,1 these several  detections are
consistent with a single extended emitting region suffering, at different 
positions, different degrees of extinction.

\subsection{Four ISOCAM bandpasses}

Fig.\,2 presents the 6.7$\mu$m, 7.7$\mu$m, 12$\mu$m and 14.3\,$\mu$m
ISOCAM maps.  The 7.7\,$\mu$m  map essentially shows  the galaxy in the
emission of a spectral feature usually attributed to polycyclic
aromatic hydrocarbon (PAH)  emission.  The 14.3\,$\mu$m map includes contributions
from warm dust, nebular line emission such as [NeII] and PAH
feature emission. The galaxy has the
approximate shape of an inverted tuning fork with regions 1, 2, 3 
and 4 dominating the emission to different degrees at different
wavelengths.  IRAS flux  densities for the system are given in
Table 3 for reference.

\begin{figure}
\resizebox{\columnwidth}{!}{\includegraphics*{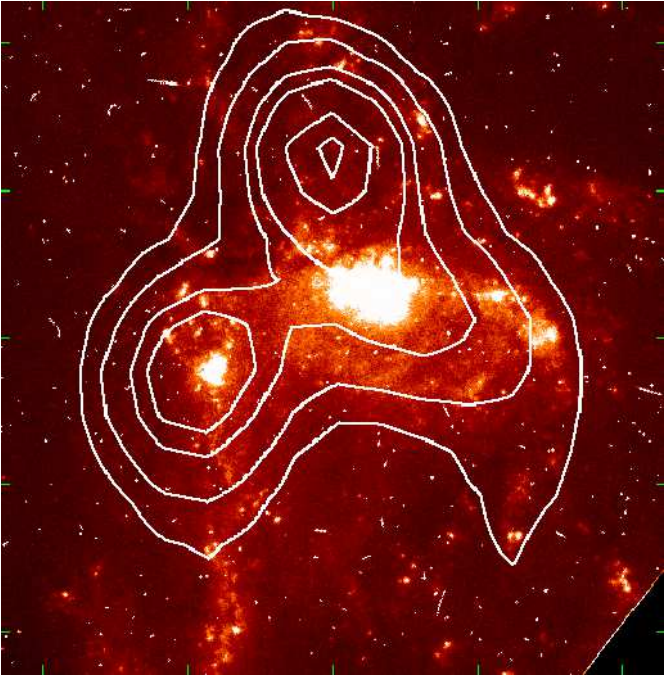}} 
\caption{The 14.3/6.7 um ratio map overlaid on the HST image of Mkn\,297.  Deconvolved
6.7 and 14.3\,$\mu$m maps  were used and the ratio  map was
convolved with the 6.7\,$\mu$m  point spread function.   The ratio
contour  levels in ascending order of brightness are: 1.0, 1.5,
2.0, 2.5, 3.0, 3.3.
}
\end{figure}

The 14.3/6.7\,$\mu$m ratio map presented in Fig.\,4 shows values
greater than unity across much of the map, which is indicative of
active star formation over a wide area.   The 14.3/6.7\,$\mu$m ratio
generally decreases as interactions develop and starbursts age because PAHs,
which dominate the shorter--wavelength band,
are  no longer  destroyed  by  the highly  ionizing  O-stars, and  dust
emission declines \cite{iso-vigroux-1999esa,iso-cesarsky-1999,iso-charmandaris-1999asp,iso-helou-1999esa}.
The ratio is highest for source regions 1, 2, 4 and 3 respectively,
with a peak  value of 3.4 at source  1 and a global value  of 1.5\ to\ 2
over much of the galaxy.


\subsection{The PHT-S Spectrum}

The  PHT-S spectrum is  presented in Fig.\,5, and line-fluxes
derived for the features in the spectrum are given in Table 4.   To
allow for the limited (vignetted) aperture size of PHT-S with respect to  the
extent of the galaxy (see Fig.\,1), the PHT-S signal was 
normalised to the ISOCAM 7.7 micron flux density as described in 
Sec.\,2.4. 

\begin{figure*}
\epsfig{file=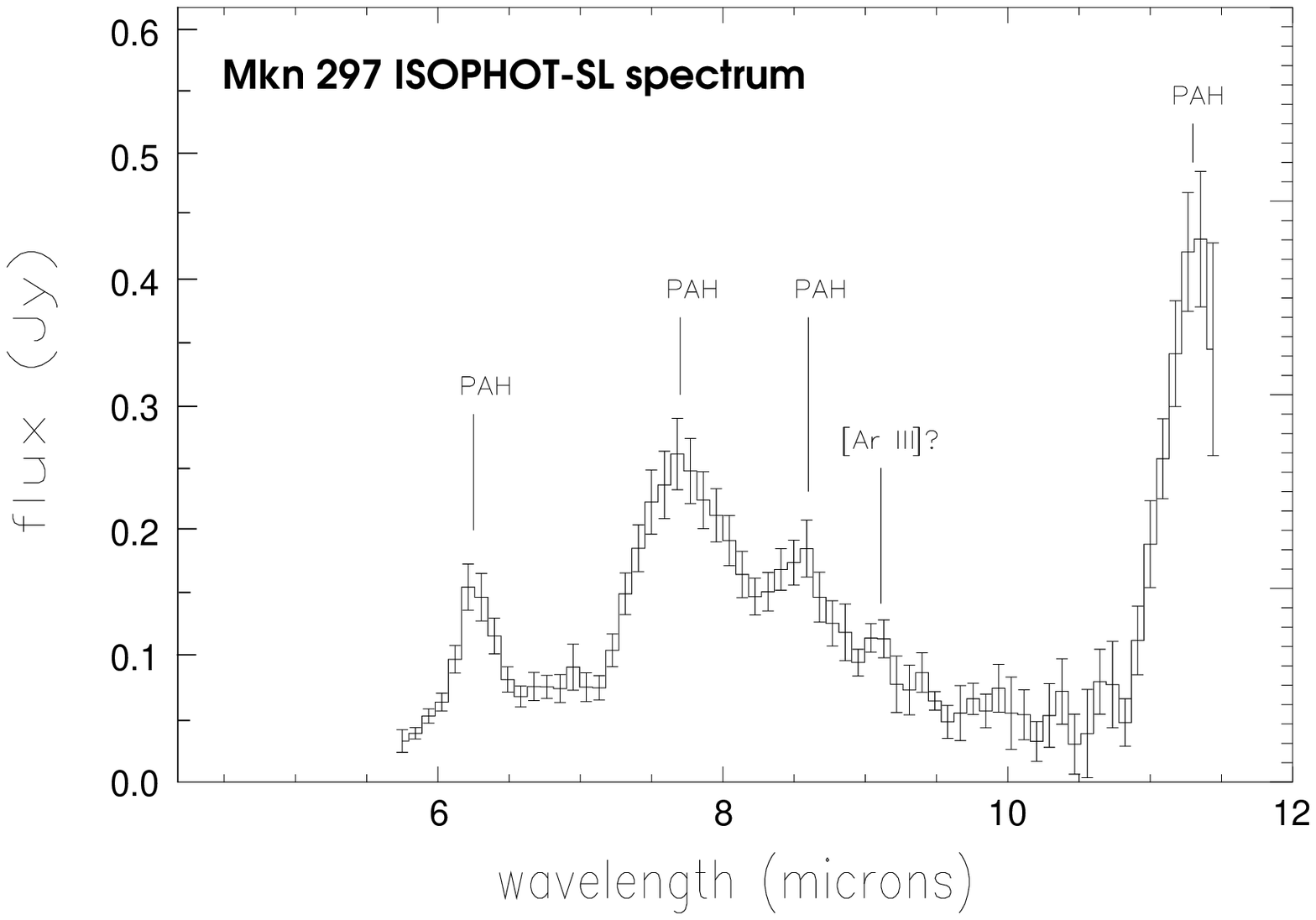,width=\textwidth,clip=,bb=17 320 593 773}
\caption{PHT-SL spectrum of Mkn\,297, with the PHT-S aperture
centred 3$''$ East of the nucleus of galaxy A. The PAH features are indicated. 
The spectrum has been Hanning smoothed. The flux levels are those recorded in the PHT-S aperture,
before scaling to the larger ISOCAM field-of-view (see Sec.\,2.4).}
\end{figure*}

The most marked characteristic of the PHT-S spectrum is the
presence of the  family of infrared bands at 6.2, 7.7, 8.6  and
11.3\,$\mu$m that  are generally attributed  to emission from 
PAHs.  There is a possible feature, seen at about the 3-sigma level, corresponding to
the nebular line emission from [Ar III] at a rest-frame wavelength 8.99\,$\mu$m. There is  
no convincing  evidence of [S IV] at 10.51\,$\mu$m.  These features have been
observed in other starburst galaxies (Metcalfe et al.\,1996, Helou et al.\,1999,
Laureijs et al.\,-2000, O'Halloran et al.\,2000, 2002 and 2005).

\begin{table*} 
\caption[]{PHT-SL fluxes for Mkn\,297. The columns give, respectively, 
the line identification, the  wavelength range, and the integrated line-flux and
the integrated line-flux scaled to the full ISOCAM footprint, as described in
Sec.\,2.4.
}
\begin{center}
\begin{tabular}{lccc} 
\hline\noalign{\smallskip}  
Line  ID   &  Wavelength range  &   Flux                                   &  Flux                                     \\
($\mu$m)   &   ($\mu$m)         &  ($\times$10$^{-15}$ W/m\(^{2}\) )       &  ($\times$10$^{-15}$ W/m\(^{2}\) )        \\
           &                    & (in the PHT-S $24''\times24''$ aperture) &  scaled to ISOCAM 7.7$\mu$m beam          \\
\noalign{\smallskip}
\hline
\noalign{\smallskip}
PAH 6.2    &  5.8 - 6.6         &   3.0\,$\pm$\,0.8                        &       4.4\,$\pm$\,1.2                     \\
PAH 7.7    &  7.2 - 8.2         &   9.2\,$\pm$\,2.4                        &      13.4\,$\pm$\,3.6                     \\
PAH 8.6    &  8.3 - 8.9         &   2.7\,$\pm$\,0.6                        &       3.9\,$\pm$\,0.9                     \\
PAH 11.3   & 10.9 - 11.6        &   1.1\,$\pm$\,0.3                        &       1.7\,$\pm$\,0.5                     \\
\noalign{\smallskip} 
\hline
\end{tabular} 
\end{center}
\label{isocam-fluxes}
\end{table*}

\subsection{The LWS Spectrum}

\begin{figure} 
\resizebox{\columnwidth}{!}{\includegraphics*{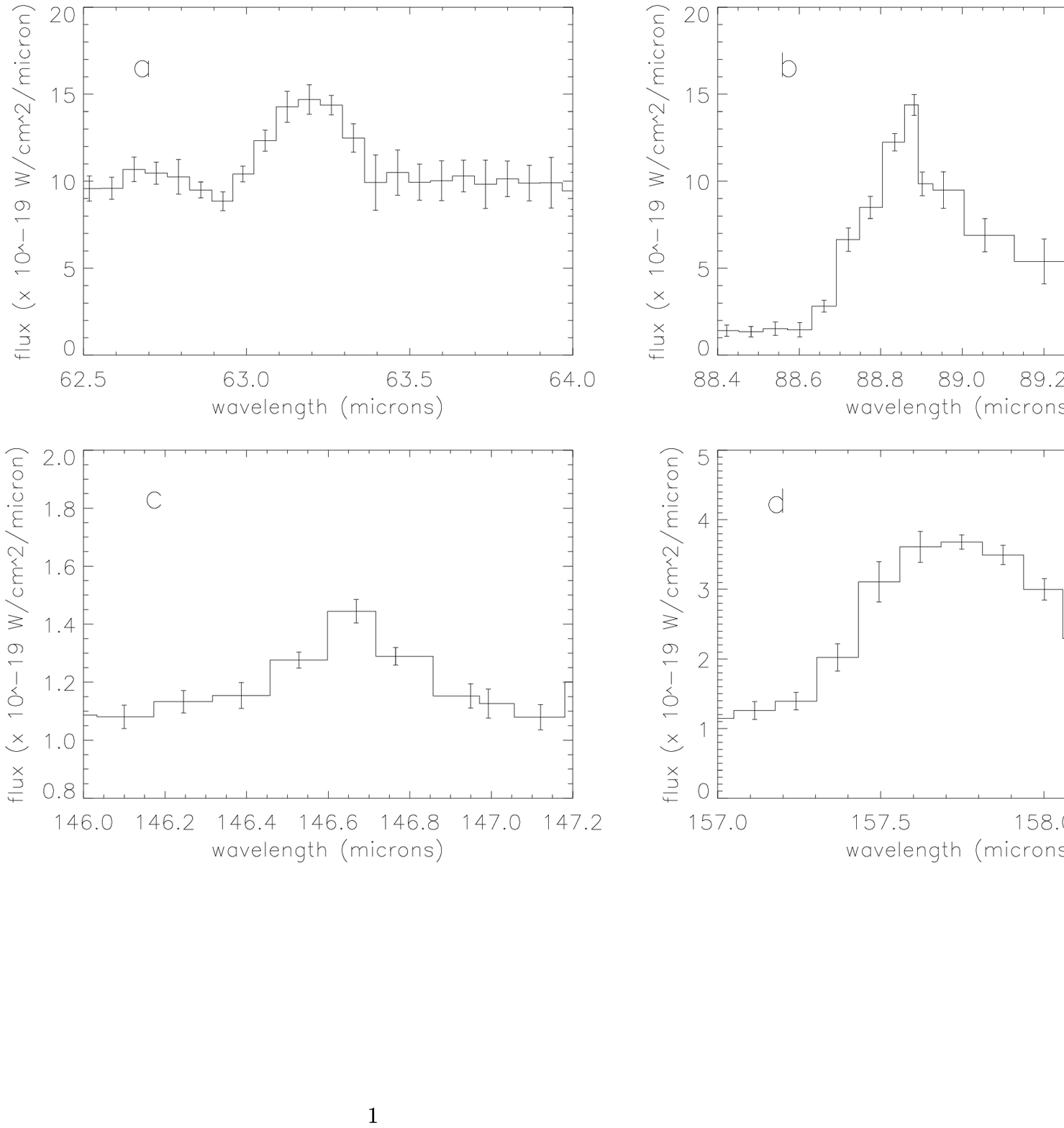}}
\caption{LWS spectra of (a) the [OI] line at 63.1\,$\mu$m, (b) the [OIII]
line at 88.8\,$\mu$m  (top right), (c) the [OI] line  at 146.7\,$\mu$m and
(d) the [CII] at 157.7\,$\mu$m.}  
\end{figure}

The LWS line fluxes for the far-infrared lines at 157.7\,$\mu$m [CII], 
88.8\,$\mu$m [O II] and for [OI] at 63.1\,$\mu$m and
146.7\,$\mu$m are listed in Table 5 and corresponding spectra are
presented in Fig.\,6.

\begin{table}
\caption[]{LWS line fluxes  for Mkn\,297. The  three columns
give the  line identification, the wavelength and the line-flux
respectively.}
\begin{center}
\begin{tabular}{lcc}
\hline\noalign{\smallskip}  
 Line        &  $\lambda$ &    Line-Flux                        \\
             &  ($\mu$m)  &  ($\times$10$^{-15}$ W/m\(^{2}\) )  \\
\noalign{\smallskip}
\hline 
\noalign{\smallskip}  
$[$O I$]$    &     63.1   &     2.0\,$\pm$\,0.5                 \\
$[$O III$]$  &     88.8   &     1.8\,$\pm$\,0.4                 \\  
$[$O I$]$    &    146.7   &     0.17\,$\pm$\,0.07               \\  
$[$C II$]$   &    157.7   &     1.8\,$\pm$\,0.1                 \\
\noalign{\smallskip}
\hline
\end{tabular}
\end{center}
\end{table}

Table 6 lists the continuum source strength recorded in the ten
LWS  detectors by binning data across the spectral range of each
detector. The spectral energy distribution of Mkn\,297 using ISOCAM,
PHT-SL, LWS and IRAS fluxes is presented in Fig.\,7.

\begin{table}
\caption{Flux densities measured with LWS, binned to yield a single
number per LWS detector. 
}
\begin{center}
\begin{tabular}{ccc}\hline
\noalign{\smallskip}
Detector label  &   $\lambda_\mathrm{ref}$ & Flux Density \\
                &  ($\mu$m)                &     (Jy)     \\
\noalign{\smallskip}
\hline
\noalign{\smallskip}
      SW1        &  46.8                   &     2.5      \\
      SW2        &  56.0                   &     4.7      \\
      SW3        &  63.7                   &     6.2      \\
      SW4        &  74.4                   &     8.4      \\
      SW5        &  83.3                   &     8.8      \\
      LW1        &  97.2                   &     10.4     \\
      LW2        & 115.7                   &     9.9      \\
      LW3        & 137.8                   &     9.1      \\
      LW4        & 156.9                   &     8.5      \\
      lW5        & 175.2                   &     7.2      \\
\noalign{\smallskip}\hline
\end{tabular}
\end{center}
\label{lws-broad}
\end{table}

\section{Discussion}

\subsection{Model of the interacting system}

Taniguchi   \&  Noguchi  \cite*{tn:1991},   using  numerical   N  body
simulations,  were able  to account  for the  morphology  and velocity
structure  of  Mkn\,297.  The  simulations  suggested  that a  coplanar
radial  penetration  between  two   disk  galaxies  yielded  the  wing
formation which is seen now, about 1.5\,$\times$\,10$^{8}$ years after 
the collision, when the  disk  of an  edge-on (from the Earth's perspective) 
galaxy has been deformed  into a  wing structure.  ISOCAM source 2 (Fig.\,1) 
corresponds to the nucleus of the edge-on  galaxy, and the North-South  elongated  
structure corresponds to poorly
formed tails emerging from  the nucleus of this galaxy \cite{tn:1991}.
One of the best examples of a galaxy exhibiting such a tail morphology
is the Antennae (NGC\,4038/4039), where the two well developed
tails  are considered  tidal debris  from  each component galaxy.   The tail  of
Mkn\,297 differs from the Antennae because it is not so well developed
and  is more  luminous in  the infrared, with a  high 14.3/6.7\,$\mu$m
ratio indicative of massive star formation.

The future evolution of the two galaxies has been predicted based on the
simulations.  After the wing phase,  the stellar disk is deformed into
a bar-like structure, gas clouds tend to sink into the nuclear regions
of the edge-on galaxy and  the nuclear starburst becomes dominant.  It
is  expected  that Mkn\,297  will  evolve  into  two galaxies,  one  a
starburst nucleus galaxy and the other a ring galaxy. (See also 
Efstathiou et al.\,2000.)

The simulations also suggest a separation between the two galaxies of
about  four  galactic  radii,  and  hence the  orbital  plane  of  the
collision must be significantly inclined from the plane of the sky (by
about 70  degrees, to  account for the  apparent closeness of  the two
galactic nuclei).

\subsection{HII and CO observations}

Mkn\,297  has an optical spectrum typical of a HII region, with
line  intensities  consistent with excitation by massive  young
stars.  The nuclear region  of galaxy A  (source  4 in Fig.\,1) has
the  highest optical surface brightness, with a strong  stellar
continuum  and an  abundance of  [O/H] $\approx$\,8.7, which  is 
near the  solar  value  and  typical of late-type spiral galaxies
\cite{lls:1992}.  The nuclear region of galaxy B (source 2 in
Fig.\,1) has a much bluer spectrum,  with an excitation 
comparable  to  that  of  the HII  regions in  the  Large
Magellanic Cloud \cite{bur:1988}. 

Compact  CO  line  emission  from  Mkn\,297 has  been  mapped  and, at 
the resolution of the systems used (of the order of 20$^{\prime\prime}$\,HPBW) has
been associated  with  the  optical knots  \cite{sls:1993,sthw:1990}.  The
estimated  mass of  H$_{2}$  is $\sim\,8\,\times\,10^{9}$\,M$_{\odot}$
and    is    approximately   75\%    of    the    HI   mass    of
1.1\,$\times$\,10$^{10}$\,M$_{\odot}$.

\subsection{Far infrared spectral analysis}

Atomic oxygen  and ionized  carbon are the  principal coolants  of the
gaseous interstellar medium via their  fine structure lines in the far
infrared  (FIR).   In particular,  [OI]  at  63\,$\mu$m  and [CII]  at
158\,$\mu$m  dominate  the cooling  in  the photodissociation  regions
associated with massive stars such  as Wolf Rayets, along with [OI] at
146\,$\mu$m  and [OIII] at  88\,$\mu$m \cite{san:2003}.   The [OI]
and [CII]  features are  also produced in  the warm atomic  gas behind
dissociative shocks,  in HII  regions or in  photodissociation regions
(PDRs),  while  [OIII] is  more  associated  with denser  environments
within HII regions \cite{malhotra1997,braine1999}.

The [OI] 63\,$\mu$m and [CII] 158\,$\mu$m lines were well detected in Mkn\,297
and were used to determine the luminosity ratios
L$_{CII}$/L$_{FIR}$    and (L$_{CII}$+L$_{OI}$)/L$_{FIR}$,
which probe the nature of  the environment within
the galaxy.   Values of  L$_{CII}$/L$_{FIR}$ = $5.1
\times 10^{-3}$    and (L$_{CII}$+L$_{OI}$)/L$_{FIR}$ = $10.7 \times
10^{-3}$   were determined using the IRAS flux-densities \cite{malhotra1997}.
The values found are 
consistent with those from  other starburst galaxies  
\cite{malhotra1997,braine1999},  though  for   higher  dust
temperatures and star formation  rates these ratios decrease \cite{malhotra1997}.
Mkn\,297   falls  at  the  higher  star
formation/dust     temperature     end     of     this    
correlation, which provides further evidence of strong ongoing star
formation within this galaxy.

The [CII] line emission is the dominant gas coolant in most regions
of atomic interstellar gas and therefore reflects the heating input
to the gas (Malhotra et al. 2001).  The ratio of [CII]/far-infrared
(FIR), as a function of the ratio of the flux density at 60\,$\mu$mµm to
100\,$\mu$m, R(60/100), has been measured for a large sample of galaxies
and used to study the radiation field (Helou et al. 2001).  
The
[CII]/FIR ratio for Mkn\,297 is 0.51$\times\,10^{-2}$ and R(60/100) is 0.64 and
these values are typical of a star forming galaxy as shown in
Figure 1 in Helou et al. (2001).  A strong correlation was also
found between [CII] and the integrated mid-infrared flux in the range 5-10\,$\mu$m
in a large sample of star forming galaxies.  The mid-infrared flux
is dominated by aromatic features (AFEs), that are generally
associated with the smallest interstellar grains.  The ratio of the
two quantities [CII]/AFE, where AFE is $\nu f(\nu)$ over the wavelength
range 5-10$\mu$m, is nearly constant at 1.5\% over a wide range in
values of R\,(60/100) (Helou et al. 2001, Dale et al. 2000).  
In Mkn\,297, the value of $\nu f(\nu)$ was
synthesised from the PHT-SL spectrum (Fig.\,5), extrapolating the
continuum to 5 microns in the integral.  The resulting value was scaled to the
area of the ISOCAM footprint, as described in Sec.\,2.4, resulting in a high value
of [CII]/AFE, 4\%.  
Although the value of 4\% might be regarded as an upper limit because the PHT-SL 
measurement covered a much smaller field of view than the LWS spectrometer 
(Table 1). However, as argued in Sec.\,2.4, it is unlikely that much flux in 
the PHT-SL range falls outside of the ISOCAM footprint to which it has been scaled. 
The value of 4\% is an additional argument in favour of star formation
dominating the mid-infrared emission.

The unidentified infrared bands (UIBs) dominate the mid-infrared
emission from Mkn\,297 (Fig.\,5) with little or no emission in the
wavelength region between 9 and 10\,$\mu$m from very small grains
(VSGs).  This result is also observed in IC 694 (Source A in Arp
299) but not in knot A in the Antennae.  In cases where the
mid-infrared emission is dominated by UIBs, the SFR can be obtained
from the following relationship (Roussel et al. 2001) :

SFR (M$_{\odot}$\,yr$^{-1}) = 6.5\,\times\,10^{-9}\,L(15\,\mu)$

The flux density at 15\,$\mu$m was converted into luminosity using a
bandpass of 6.75\,THz. This yields SFR in Mkn\,297 of 17.1\,M$_{\odot}$\,yr$^{-1}$.  
Taniguchi and Noguchi (1991) state a FIR-based value of 23.5\,M$_{\odot}$\,yr$^{-1}$
for the SFR in Mkn\,297.

The SFR can also be obtained from the $H_{\alpha}$ luminosity
using (Lee et al.\,2002):

SFR (M$_{\odot}$\,yr$^{-1}$) = $7.9\,\times\,10^{-42}\,L(H_{\alpha})$\,ergs\,s$^{-1}$

The H$_{\alpha}$ flux from Mkn\,297 is 21\,$\times\,10^{-13}$\,ergs\,s$^{-1}$\,cm$^{-2}$ (Deeg
et al. 1997) which yields a value of 8.6\,M$_{\odot}$\,yr$^{-1}$.  These 
results suggests that about half of the star formation in Mkn\,297 is
hidden in the optical.

\subsection{Dust components within Mkn\,297 and the FIR-radio correlation}

A dust model for Mkn\,297 is shown in Fig.\,7, 
denoted by the solid curve. It includes two separate dust populations: 
a warm dust component at 130\,K and a cooler dust component at 
38\,K \cite{iso-krugel-1994aa,iso-siebenmorgen-1999aa}.

There is growing  evidence for the existence of several components in
the dust  distribution of galaxies \cite{kla:2001,pope:2002} that  can be broken
down into  warm dust components  with T\,$\geq$\,20\,K  associated with
star formation regions, and  a spatially extended distribution of cold
dust with T\,$\leq$\,20\,K.  

Very small dust grains and PAHs are transiently heated by the
single-photon emission process to temperatures much higher than
40\,K, up to several hundred degrees Kelvin \cite{des:1990}, out of
thermal  equilibrium with their environment \cite{cal:2000}. These
account for the bulk of the emission in the PHT-S waveband and for 
emission at wavelengths shorter than 40\,$\mu$m.  The ``warmer" or
``hot dust" component, the 130\,K component in Fig.\,7, is due to small grains. 

Dust emission from HII regions is dominated by large dust grains heated
by the intense radiation field. These grains can reach temperatures in 
excess of 20\,K up to more than 40\,K. Due to the nature of the blackbody, 
this emission will outshine the cold dust in these regions. The 38\,K dust 
component in Fig.\,7 comes from large grains in HII regions. 

The  coldest   dust  emission  is  associated  with large  
grains  emitting  at  wavelengths   in  excess  of 80\,$\mu$m  
and in   thermal   equilibrium   with    their   environment
\cite{cal:2000}.  A small contribution  at 100\,$\mu$m is also made
by very small  dust grains. The large dust  grains  account 
for   practically all  the emission longward of  80\,$\mu$m in
galaxies, with the  maximum of the SED occurring close to
200\,$\mu$m. We have not detected this very cold component in  Mkn\,297
within the sensitivity of the current measurements (and given
the occurrence of cirrus and zodiacal light confusion at the
longest wavelengths).

\begin{figure}
\resizebox{\columnwidth}{!}{\includegraphics*{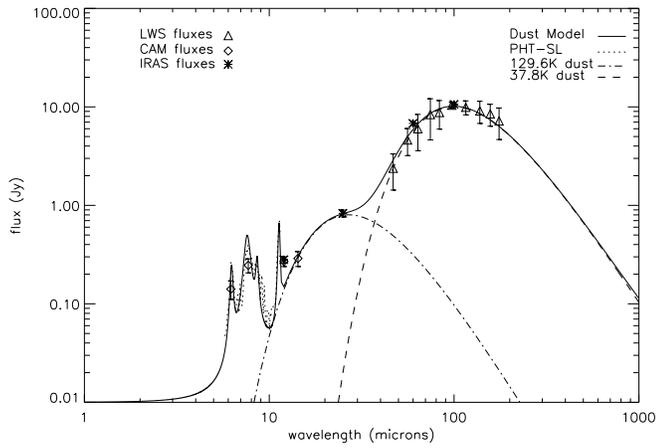}} 
\caption{Spectral energy  distribution of Mkn\,297  using
ISO and IRAS flux data, including  the key for the different
symbols.  The PHT-S spectrum has been normalised to the CAM
7.7\,$\mu$m data point as described in the text. The
models fitted to the data are described in Sec.\,4.4.}  
\end{figure}

The luminosity of each component was found by integrating over a
model greybody, fitting temperatures and calculating a
dust mass for the component \cite{kla:1991}.  Using
this method, the IR luminosities and dust component masses  in 
similar  starburst  galaxies  such  as  the  Antennae  and
NGC\,6240  \cite{kla:1997}  were  calculated  in order  to  check 
the validity of the derived values.  The values obtained were
generally in good agreement 
The dust model yields for Mkn\,297 a total L$_{IR}$ of 9.7 x  10\(^{10}\) L$_{\odot}$.  
For each component, the luminosities and  dust masses are given in 
Table 7. The  warmer dust emission is concentrated within
the  disk and associated with the strongly ionizing sources found
in star forming regions.

%
%

\begin{table}
\caption[]{Total infrared luminosity and dust component masses for Mkn\,297. The total FIR 
luminosity of 9.7$\times\,10^{10}$\,L$_{\odot}$ places this galaxy right on the
threshold of LIRG status.}
\begin{center}
\begin{tabular}{lcc}
\hline\noalign{\smallskip}
 Dust component &   L$_{IR}$                          &         Mass              \\
                &  [L$_{\odot}$]                      &     [M$_{\odot}$]         \\
\noalign{\smallskip}
\hline
\noalign{\smallskip}
     130\,K     &   7.5\,$\pm\,0.8\times10^{9}$   &  4.2\,$\pm\,0.4\times10^{3}$  \\
      38\,K     &   8.9\,$\pm\,0.9\times10^{10}$  &  3.3\,$\pm\,0.4\times10^{7}$  \\
\noalign{\smallskip}
\hline
\end{tabular}
\end{center}
\end{table}

Deeg et al. (1997) report radio data for a source within 3$^{\prime\prime}$ of
Source 1 having flux densities of 244\,$\pm$\,20\,mJy and 
104\,$\pm$\,7\,mJy at 0.325\,GHz and 1.489\,GHz respectively.
This is consistent with the variable radio source reported by 
Heeschen et al. (1983), which varied at 4.885\,GHz\,(6cm) from 4.2\,mJy on January 12 1980
to 12.2\,mJy on April 23 of that year. See also Yin et al.\,(1994), and Fig.\,8.

Pierini et al. \cite*{pi:2003} give expressions to relate warm and cold FIR emission 
to radio flux, and the values for FIR emission for Mkn\,297 given in Table 7 fall 
on the correlations in each case. From Condon et al.\,(1990), 
log\,(L(1.4 Ghz))\,=\,22.74, while Table\,7 yields log\,(L(FIR))\,=\,37.85.
For the warm and cold dust respectively:
log\,(L(FIR))\,=\,36.74 and log\,(L(FIR))\,=\,37.82.
All lie close to the relevant correlations, though it must be remarked that
the total FIR luminosity for Mkn\,297, and particularly the cold component, fall towards
the high FIR extreme of the Pierini et al. correlations.

\subsection{The hidden source in Mkn\,297 and comparison with other cases}

The JHK 2MASS images of Mkn\,297 reveal a source that is North of
galaxy A and coincides with the star forming regions 7 and 8 of
Hecquet et al.\,(1987). An overlay of the ISO 15\,$\mu$m map and
the 2MASS K-band map on the HST image is shown in Fig.\,3. The
K-band peak clearly corresponds to the optically evident HII region,
and not  at all to the hidden MIR source - Source 1.  The star
forming region  that is nearest to the hidden source is region 16
of Hecquet et al.\,(1987) and has the  largest extinction of all
the star forming regions in Mkn\,297.  The hidden source in Mkn\,297
is inconspicuous in the optical and also in the JHK 2MASS images.  

There are three other sources where strong mid-infrared emission
has been detected that are unremarkable in the optical.  The first
source is knot A in the overlap region of the colliding galaxies in
the Antennae (Vigroux et al. 1996, Mirabel et al. 1998), the second
is in the intragroup region of Stephan's Quintet that could be an
intruding galaxy (Xu et al. 1999) and the third is source A or
IC694 in the interacting galaxies in Arp 299 (Charmandaris et al.
2002, Gallais et al. 2004).  The hidden source in Mkn\,297 is
important in this context and is very similar to those in the
Antennae and Stephan's Quintet because the hidden sources do not
coincide with the nuclei of any of the interacting galaxies.

At 15\,$\mu$m (LW3), the hidden source in Mkn\,297 is 14.6 time more 
luminous than knot A in the Antennae, and 3.8 time more luminous than 
the hidden source in Stephan's Quintet. However, its luminosity is 
only 0.09 times that of the hidden source IC 694 in Arp 299. However 
the ratio of the LW3 flux from the hidden source to the total flux is
similar for all sources with values of 0.23, 0.19, 0.26 and 0.10
for Mkn\,297, the Antennae, Arp 299 and Stephan's Quintet
respectively.

The hidden source in Mkn\,297 appears to be more similar in properties 
to knot A in the Antennae, than it is to the nuclear source IC
694 (Charmandaris et al. 2002, Gallais et al. 2004).  The source in
IC 694 becomes visible as a point source in the JHK 2MASS images
and gets brighter from J to H to K and also contributes
significantly to the near infrared radiation from the galaxy. 
However this is not the case in Mkn\,297 or the Antennae, where knot A
is inconspicuous in the near infrared and becomes dominant in the
mid-infrared.  Table 8 lists comparative infrared properties for
Mkn\,297 and three sources with somewhat similar morphological 
characteristics.

\begin{table} 

\caption{A comparison of the properties the hidden sources (h.s) in four systems.
Column 3 lists the total 15\,$\mu$m emission from the host galaxy in each case.}
\begin{center} 
\begin{tabular}{lcccccc}
\hline 
\noalign{\smallskip}
  Source   &  7\,$\mu$m & 15\,$\mu$m  & 15\,$\mu$m  &  D   &  R1  &  R2 \\
           &    mJy     &    mJy      &    mJy      & Mpc  &      &     \\
           &   (h.s.)   &   (h.s.)    &   (host)    &      &      &     \\
\noalign{\smallskip}
\hline
\noalign{\smallskip}
Mkn\,297   &   42.5     &  66.5       &    289      & 66   &  -   & 0.23\\
Antennae   &   19.1     &  49.7       &    261.7    & 20.0 & 14.6 & 0.19\\
Stepn's Q. &    -       &  11.9       &    114.9    & 80   &  3.8 & 0.10\\
Arp 299    &   325      &  1860       &    7027     & 41   & 0.09 & 0.26\\
\noalign{\smallskip} \hline
\noalign{\smallskip} 
\multicolumn{7}{l}{  R1 = Ratio of 15\,$\mu$m (LW3) emission from the hidden }     \\
\multicolumn{7}{l}{  source in Mkn\,297 to the hidden source in the other galaxy } \\ 
\multicolumn{7}{l}{  (corrected for distance to Mkn\,297.)}                        \\ 
\multicolumn{7}{l}{  R2 = Ratio of 15\,$\mu$m emission from each hidden source }   \\ 
\multicolumn{7}{l}{  to the total 15\,$\mu$m emission for its host galaxy. }                                                                                                                          \\
\noalign{\smallskip} 
\noalign{\smallskip}
\end{tabular} 
\end{center} 
\label{many-galaxies}
\end{table}

The differences between the hidden sources suggest that dust
extinction is lower in the hidden source in IC 694 than in Mkn\,297 or
knot A in the Antennae.  An alternative interpretation is that the
thermal emission from an enshrouded AGN becomes visible in the near
infrared.  This phenomenon was observed in the nucleus of NGC 1068
(Alonso-Herrero et al. 2003) and will be discussed further in
Sec.\,4.6.  The variable and compact radio source Mkn\,297A (Fig.\,8) that
falls on the edge of the hidden source in Fig.\,1 has been studied 
extensively (Hummel et al.\,1987, Condon et al.\,1991, Yin and 
Heechen 1991, Yin 1994).  This source
was unresolved with the resolutions of the VLA and the flux density
shows a rapid rise followed by a slower decline with maximum flux
density occurring first at the shortest wavelength (Yin 1994).  The
flux density, S, was modelled as a power law function of both
frequency and time of the form 

$S\,\propto\,\nu^{\alpha}\,t^{\beta}$

which is characteristic of radio supernovae.  The values of the
indices $\alpha$ and $\beta$ for the variable radio source are typical of radio
supernovae.  These results reveal that the radio source close to the hidden source 
is a radio supernova and not a dominant AGN.

The VLA map of Mkn\,297 (Fig.\,8) shows that the
variable radio source coincides with star forming region 14 of Hecquet et al.\,(1987).

\subsection{Does Mkn\,297 harbour an AGN?} 

Several  diagnostic tools have been proposed to probe  the nature  of the
activity  within a central  starburst  region.  Lu et al.\,\cite*{iso-lu-1999esa} 
cite the  ratio of the integrated  PAH luminosity and the 40 to 120\,$\mu$m  
IR luminosity as providing a tool to discriminate between
starbursts, AGN and  normal galaxies. The lower  the ratio the
more likely the  galaxy harbours an AGN, due to  the dominance of very
small  dust grain emission  powered by the AGN in  the 40  to 120\,$\mu$m
region  \cite{iso-vigroux-1999esa}.  For Mkn\,297,  the ratio
is 0.11, which is typical of a strong starburst.  Similarly, 
Lutz et al.\,\cite*{iso-lutz-1998apj}, Genzel et al.\,\cite*{iso-genzel-1998apj} 
and Laureijs et al.\,\cite*{iso-laureijs-2000aa} 
state that the ratio of the  7.7\,$\mu$m PAH flux to the nearby continuum can 
provide a measure of  the level of star formation activity within
the nucleus, given the ubiquity of  strong PAH features in regions of
high star formation \cite{iso-clavel-2000aa}.   
The ratio  for  Mkn\,297 is 3.7. Plotting this value, and the 
5.9\,$\mu$m/60\,$\mu$m flux densities  
(0.061\,Jy in the PHT-S aperture/6.79Jy from IRAS) on Fig.3 of 
Laureijs et al.\,(2000) indicates strong  star formation,
suggesting that an AGN within Mkn\,297 is unlikely (O'Halloran et al.\,2000, 
Laureijs et al.\,2000).

Another set of diagnostics uses empirical criteria based on the
fact that mid-infrared emission from star forming or active
galaxies arises mostly from HII regions, photo-dissociation regions
(PDRs) and AGNs (Laurent et al. 2000).  The diagnostic diagrams
provide quantitative estimates of the contributions of AGN, PDR and
HII regions.  These diagrams are referred to as the Laurent
diagnostic diagrams (Peeters et al. 2004).  In Mkn\,297 the ratio 
of LW3/LW2 is 2 and that of LW2/LW4 is roughly 2, where the LW4
value was synthesised from the PHT-SL spectrum, making approximate 
allowance for the 
fact that the LW4 bandpass extends to shorter wavelength than PHT-S.  
These values show
that the dominant contribution to the mid-infrared emission is from
PDR and HII regions in Mkn\,297 and not from an AGN.

However, mid-IR diagnostics cannot probe beyond values of about 10
magnitudes of  extinction, and so miss any AGN embedded deep into a
star forming region with, say, 100 magnitudes of extinction. The
absence of an AGN  within Mkn\,297 can therefore only be
tentatively suggested until it is  confirmed by hard X-ray
observations.

\subsection{Unusual radio supernova remnant and possible SN/GRB link}

Heeschen et al. \cite*{hhy:1983}  found a  peculiar,  compact  and
variable radio source which falls about 2 kpc north of the nucleus of 
galaxy A, on the edge of the strongest infrared source 
in Fig.\,1 (source 1). They suggested, among other possibilities, that it might 
be a complex of supernova remnants.
In subsequent radio  VLA observations \cite{yh:1991,yin:1994} a
decay in flux density from 14\,mJy to 4.7\,mJy at 20\,cm was recorded 
between 1983
and 1990, and the source (Fig.\,8) was attributed to a single, very luminous, 
radio supernova that exploded in mid-1979.  The radio SN was later cataloged
as SN\,1982aa, of uncertain type because of the paucity of radio results within 
its first 8 years and the absence of optical observations \cite{gre:1994}. 
VLBI  observations  
\cite{lls:1992}  are consistent with the single supernova occurring in a
molecular cloud.  

The  most surprising aspect  of the supernova  interpretation was the  very high
radio luminosity  of $\sim$\,1.5\,$\times$\,10$^{5}$\,L$_{\odot}$.  As
a  radio  supernova  it is  one  of  the  most energetic  events  ever
observed, alongside SN\,1986\,J,  41.\,9\,+\,58 in  M82, SN\,1998bw and 
SN\,2003dh \cite{wdeb:1990,gal:1998,ber:2003}.  These powerful supernovae
are  more energetic,  by  a  factor  of  10,  than  canonical  supernovae which
release $\geq$\,10$^{52}$\,ergs during the explosion. Indeed, SN\,1982aa  
has a  total released  energy of  10$^{53\,\pm\,0.5}$\,ergs, as derived
from the radio flux reported by \cite{yin:1994}.  

So far,  many of the most powerful supernovae 
have been of Type Ib/c, the best known  example of which has been SN\,1998bw, one
of     the     most     luminous     radio    SN     ever     observed
\cite{weil:2001,weil:2002}.  Interestingly, SN\,1998bw has
been strongly associated with a weak gamma-ray burst (GRB), GRB\,980425
\cite{gal:1998,kouv:2004,sod:2004}.  It  has only  been exceeded 
in radio luminosity by SN\,1982aa \cite{yin:1994}  and SN\,2003dh,
which is identified with the cosmological burst GRB\,030329 (Berger et al. 2003). 

A well  known model  for the  progenitors  of GRBs,
linked  to powerful supernovae,  and by  extension to  regions of  massive star
formation, is the  ``collapsar''  model. In  this  model the  core of  a
massive star  collapses to a black hole, which then powers the GRB by the accretion
of an additional solar mass  \cite{mcf1999}, and signatures consistent with
spin-up and spin-down of black holes have been found in the light
profiles of GRBs \cite{mcb:2002}. 
For   SN\,1982aa, given the   amount  of  energy  released 
it is quite plausible that a GRB may have been associated with  this 
candidate obscured supernova. 
Furthermore, Mkn\,297 is morphologically similar to host galaxies of GRBs
seen in HST images, which show most GRB host galaxies to be either interacting
or obscured (Wainwright et al. 2005).
In 1979, many spacecraft had GRB detectors 
and  a search is recommended in data from the period  July through December 1979,
for a GRB consistent with the direction to Mkn\,297. It  is  also
important  to  continue radio observations  of Mkn\,297A
so  that a proper  comparison can be  made of its long term radio 
profile  with that of SN\,1998bw, SN\,2003dh and other 
supernovae (Frail et al.\,2000) with associated 
GRBs that have yet to be discovered.

\begin{figure}
\resizebox{\columnwidth}{!}{\includegraphics*{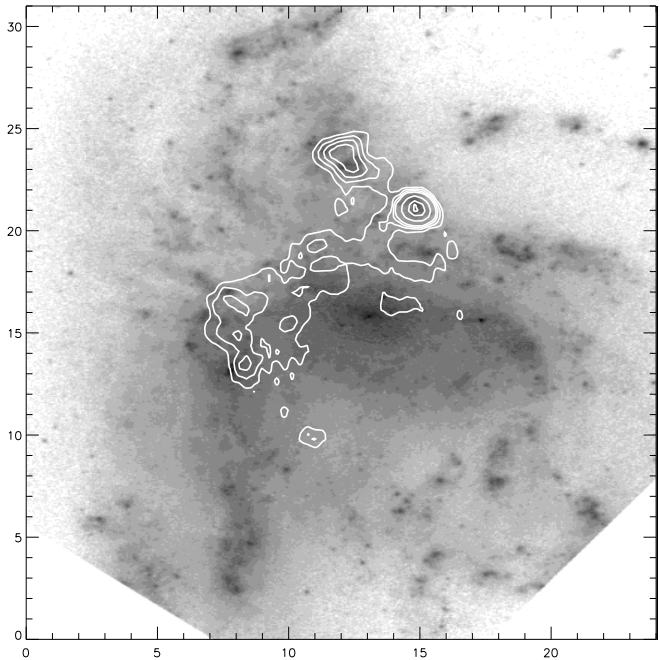}} 
\caption{An overlay of the Yin\,(1994) 6cm radio contours on the HST image of Mkn\,297.}  
\end{figure}

\section{Conclusions}
Mkn\,297 is an interacting system possibly involving a
late-type spiral and an irregular galaxy or two disk galaxies. It has 
a ``wing'' like structure extending North and South from the nuclear
region. It was observed with ISOCAM, ISOPHOT and LWS on board ISO.  The
results obtained with the four ISOCAM filters used revealed a complex
mid-infrared emission pattern with many clumps occurring within 
an overall envelope. The two nuclear regions and the wing structure were 
easily detected in all ISOCAM bandpasses employed.  However the strongest 
infrared source does not coincide with either of the nuclei of the interacting 
galaxies but is located several arcseconds to the North, nor is that source
seen in NIR (e.g. 2MASS) images.  Strong  PAH  
features were detected in the ISOPHOT  spectrum, while  [OI], [O III]  
and [C  II ] emission lines were detected with the LWS instrument.  Using 
fluxes from these 3 instruments, masses and luminosities were determined 
for two dust components  within Mkn\,297.  The total infrared luminosity of
9.7 x 10\(^{10}\) L$_{\odot}$ classifies it as a borderline LIRG.  Numerous 
criteria discriminating star formation from AGN activity point to the energy 
source in Mkn\,297 being star formation, and establish a stronger similarity 
between Mkn\,297 and the Antennae system (NGC\,4038/4039) rather than, for 
example, with less deeply obscured interaction-driven starbursts in IC 694 
or Stephan's Quintet. Only hard X-ray measurement, however, can definitively
eliminate the possibility that Mkn\,297 harbours a highly obscured AGN.
One of the strongest known radio supernovae, SN\,1982aa, falls in a HII 
region impinging on the brightest infrared source in Mkn\,297, and has been 
labeled  a  radio  hypernova, calling to mind the suggested association
between intensely star-forming systems (LIRGS and ULIRGS) and GRBs. Further 
observations  of  this  radio supernova are recommended,  as well as a 
search for  a gamma ray burst around the time of the explosion in 1979.

\begin{acknowledgements}

We would like to thank the referee, Dr. Vassilis Charmandaris, for 
detailed comments which greatly improved the paper. 
We gratefully thank Dr. Qi Feng Yin for providing us with his VLA radio 
maps for study and for the radio data reproduced in Fig.\,8. We thank 
Dr. Avishay Gal-Yam for helpful comments and suggestions.
The ISOCAM data presented in this  paper were  analysed using  CIA,
a joint  development by  the ESA  Astrophysics Division  and  the
ISOCAM Consortium.   The  ISOCAM Consortium  is  led  by  the ISOCAM 
PI,  C. Cesarsky.  The ISOPHOT data presented in this paper was
reduced using PIA, which is a joint  development by the  ESA
Astrophysics  Division and  the ISOPHOT consortium.  LIA is a joint
development of the ISO-LWS Instrument Team at Rutherford  Appleton
Laboratories  and the Infrared  Processing and Analysis Center
(IPAC). The research of M. Burgdorf was carried out at the Jet
Propulsion Laboratory, California Institute of Technology, and
sponsored by the  National Aeronautics and Space Administration.

\end{acknowledgements}

\end{document}